\documentclass[reprint,superscriptaddress,showpacs,amsmath, floatfix,amssymb,aps,prb]{revtex4-1}
\usepackage{graphicx}
\usepackage{subfigure}
\usepackage{dcolumn}
\usepackage{bm}
\usepackage{hyperref}
\usepackage{indentfirst}
\usepackage{braket}
\usepackage{amsmath}
\hypersetup{colorlinks=true, citecolor=blue, urlcolor=blue, linkcolor=blue}
\begin{document}
\title{Temperature-dependent spectral function of a Kondo impurity in an $s$-wave superconductor}
\author{Chenrong Liu}
\affiliation{Texas Center for Superconductivity, University of Houston, Houston, Texas 77204, USA}
\affiliation{Department of Physics and State Key Laboratory of Surface Physics, Fudan University, Shanghai 200433, China}
\author{Yixuan Huang}
\affiliation{Texas Center for Superconductivity, University of Houston, Houston, Texas 77204, USA}
\author{Yan Chen}
\email{yanchen99@fudan.edu.cn}
\affiliation{Department of Physics and State Key Laboratory of Surface Physics, Fudan University, Shanghai 200433, China}
\author{C. S. Ting}
\email{ting@uh.edu}
\affiliation{Texas Center for Superconductivity, University of Houston, Houston, Texas 77204, USA}

\date{\today}

\begin{abstract}
 Using the numerical renormalization group method, the effect due to a Kondo impurity in an $s$-wave superconductor is examined at finite temperature ($T$).  The $T$-behaviors of the spectral function and the magnetic moment at the impurity site are calculated.  At  $T$=0, the spin due to the impurity is in singlet state when the ratio between the Kondo temperature $T_k$ and the superconducting gap $\Delta(0)$ is larger than 0.26. Otherwise, the spin of the impurity is in a doublet state. We show that the separation of the double Yu-Shiba-Rusinov peaks in the spectral function shrinks as $T$ increases if $T_k/\Delta(0)<0.26$ while it is expanding if $T_k/\Delta(0)>0.26$  and  $\Delta(0)$ remains to be a constant.  These features could be measured by experiments and thus provide a unique way to determine whether the spin of the single Kondo impurity is in singlet or doublet state at zero temperature.
\end{abstract}
\maketitle
\textsl{Introduction ---} The quasiparticle states induced by a magnetic impurity with spin $S=1/2$ in an $s$-wave superconductor (SC) inside the BCS gap are known as the Yu-Shiba-Rusinov (YSR) states\cite{YU-LUH,shiba,Rusinov,RevModPhys.78.373,Yazdani1767}. At zero temperature ($T$), the spectral function exhibits two $\delta$ function-like peaks symmetrically located at $\pm\varepsilon$ with respect to the center of the superconducting (SC) gap.  The physics of the YSR states was also extensively studied\cite{PhysRevLett.26.428, PhysRevB.65.024414,doi:10.1143/JPSJ.70.2860,PhysRevB.85.205410} based on theories beyond the mean field approximation or the perturbation theory. A detailed investigation of the spectral function of the Kondo impurity at $T=0$ with the Kondo coupling $J$ using the numerical renormalization group (NRG) theory has been previously carried out\cite{doi:10.1143/JPSJ.62.3181}.  Recently, the YSR states of a Kondo impurity \cite{doi:10.1143/PTP.32.37} in Fe-based SC with the spin-orbit coupling were investigated by the Bogoliubov-de Gennes equations in the mean field level \cite{PhysRevB.92.174514}.  Moreover, the physics of an Anderson impurity\cite{PhysRev.124.41}  on the interface between a topological insulator (TI) and an $s$-wave SC\cite{PhysRevLett.122.087001} was also analyzed by the NRG method.
On the other hand, there exist few experimental and theoretical works for the finite-temperature spectral properties inside the SC gap.  For an Anderson impurity with an SC lead, {\v{Z}}itko\cite{PhysRevB.93.195125} calculated the spectral properties of these sub-gap states at finite $T$ using the NRG method, and their result shows that the strengths of the YSR peaks become weakened as $T$ rises.  For finite temperature Kondo resonance, Zhang et al. \cite{Zhang} detected it on an organic radical weakly coupled to an Au (111) surface by measuring the differential conductance at low temperatures which can be described by perturbation theory of the Kondo impurity model. Moreover, Ruby et al.\cite{PhysRevLett.115.087001} probed the single-electron current which passed through the bound states on the superconducting surface and analyzed the relaxation processes of this current to obtain the information of the quasiparticle transitions and lifetimes.

Due to the exchange scattering of the thermally-excited quasiparticles with the magnetic impurity, there should be nontrivial behaviors in the impurity-site spectral functions at finite $T$.  This problem has never been seriously investigated for a Kondo impurity. It is also essential to understand whether quasiparticles could completely or partially screen the spin of the magnetic impurity in the SC at finite $T$. Besides, the relationship between the spectral function and the renormalized-magnetic moment of the impurity needs to be discussed.  In this paper, we investigate the temperature-dependent spectral function and the renormalized magnetic moment at the impurity site using the NRG method. These problems so far have not been studied for the Kondo Hamiltonian.  In Section A of the Supplemental Material (SM), the energy-evolution calculation of the Kondo impurity system at zero temperature\cite{doi:10.1143/JPSJ.62.3181} as a function of $J$ (see Fig.S1) has been reproduced. If $T_k$ is the Kondo temperature and $\Delta(0)$ is the SC gap at $T=0$, the spin at the impurity site should be completely screened and is in singlet state for $T_k/\Delta(0) >0.26$ while it is in doublet state for  $T_k/\Delta(0)<0.26$. Our calculation of the magnetic moment due to the impurity at moderate values of $T_k$ indicates that the spin of the single Kondo impurity could only be partially screened by quasiparticles for $T/\Delta(0)>10^{-2}$. The experimental consequence of our $T$-dependent spectral function will be addressed.

\textsl{Model and method ---} We consider the single Kondo impurity\cite{doi:10.1143/PTP.32.37} in an $s$-wave superconductor,
\begin{eqnarray} \label{eq1}
H=&&H_{BCS}+H_{imp}, \nonumber \\
H_{BCS}=&&\displaystyle{\sum_{k\sigma}}\epsilon_kc_{k\sigma}^\dag c_{k\sigma}-\Delta(T) \displaystyle{\sum_k}(c_{k\uparrow}^\dag c_{-k\downarrow}^\dag+H.c.), \\
H_{imp}=&&J\bm{S}\cdot \left(\frac{1}{2N_s} \displaystyle{\sum_{kk^\prime \sigma \sigma^\prime}}c_{k\sigma}^\dag \bm{\tau}_{\sigma \sigma^\prime}c_{k^\prime \sigma^\prime}\right) . \nonumber
\end{eqnarray}
$c_{k\sigma}$ is the electron annihilation operator at momentum $k$ and spin $\sigma$ ,  $\epsilon_{k}$ is the single particle energy band dispersion, $\Delta(T)$ is the $T$-dependent BCS gap parameter,  $J$ is the antiferromagnetic exchange interaction between the Kondo impurity and the conduction electrons, $\bm{S}$ is the impurity spin with $S=1/2$,  $N_s$ is the number of lattice sites, and $\bm{\tau}$ is the Pauli matrix.

Suppose that the bandwidth of the conduction electrons is from $-D$ to $D$, and the density of states (DOS) $\rho$ of the conduction electrons is taken as $\rho=1/2D$. For $\Delta(T)=0$ case, the Kondo temperature in weak-coupling limit is\cite{RevModPhys.47.773,RevModPhys.55.331,doi:10.1143/JPSJ.61.3239,doi:10.1143/JPSJ.62.3181}:
\begin{eqnarray} \label{eq2}
T_k=D(J\rho)^{\frac{1}{2}}\exp(-\frac{1}{J\rho}),
\end{eqnarray}
This result is based on the Kondo model\cite{doi:10.1143/PTP.32.37}. The ground state of the spin at the impurity site is completely screened by a conduction electron and becomes singlet at $T=0$ regardless of the magnitude of $J$. The Kondo impurity behaves like a nonmagnetic
impurity\cite{PhysicsPhysiqueFizika.3.17}.  It needs to be pointed out that there is another Kondo temperature of $T_k^*$ \cite{PhysRevLett.85.1504}  defined as the half-width at half-maximum (HWHM) of the Kondo resonance at $T=0$. To compare our results with those of others, we use both  $T_k^*$ and  $T_k$.  In Fig.S3 of the Section B of the SM, we compare  $T_k$ and $T_k^*$ as functions of $J$.

In order to carry out  the NRG method, one needs to  apply the spherical wave representation,  and to discretize the states of conduction electrons in a logarithmic way. Eq.(\ref{eq1})  is then transformed into a one-dimensional Wilson chain\cite{doi:10.1143/JPSJ.61.3239,doi:10.1143/JPSJ.62.3181}. Its brief description is given at the beginning of Section B in the SM.   One efficient way to optimize the calculation is to set $\Lambda=2$ and $N_z=8$ (the interleaved discretization grids($z-averaging$))\cite{RevModPhys.47.773,PhysRevB.33.7871,PhysRevB.49.11986,PhysRevB.79.085106,PhysRevB.93.195125}.  Furthermore, we fixed $\Delta(0)/D=0.01$  and varied $J$ at $T=0$. We also employ finite temperature SC gap $\Delta(T)$ to perform the calculation of the  $T$-dependent spectral function(See the details in Section B of the SM).  We kept at least 5000 states for the spectrum function calculations.
\begin{figure}
\centering
\includegraphics[scale=0.9]{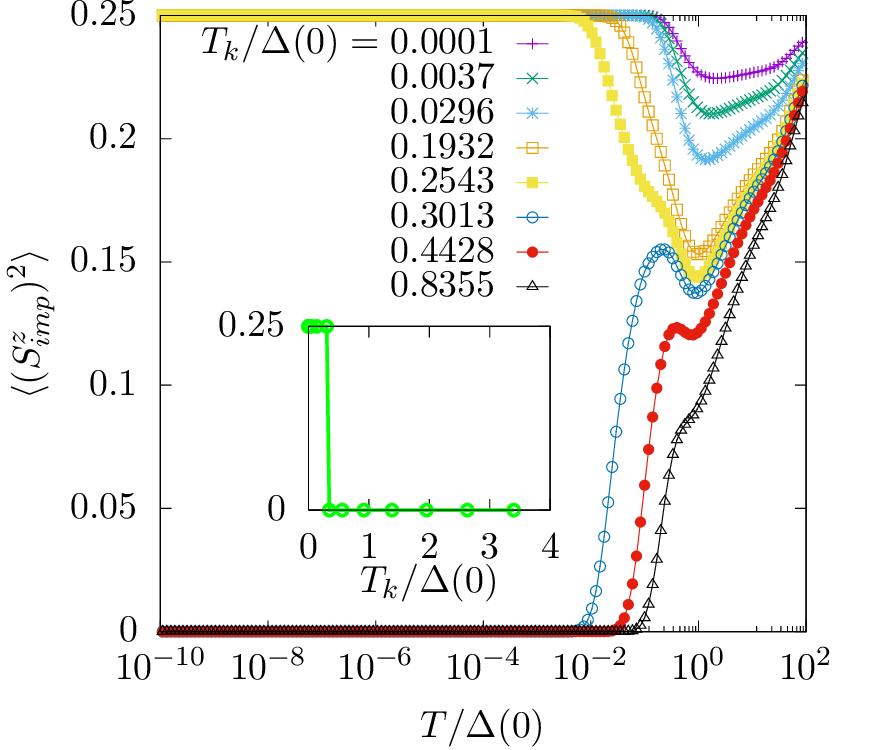}
\caption{\label{Fig1}(Color online)  The square of the impurity magnetization $\langle(S_{imp}^z)^2\rangle$ vs $T$ for different values of $T_k/\Delta(0)$. The insert figure is the impurity magnetic moment square $\langle(S_{imp}^z)^2\rangle$  at zero temperature. The crossing point is $T_k/\Delta(0) \approx 0.26$(or $J/D=0.39$).}
\end{figure}
\begin{figure*}
\centering
\includegraphics{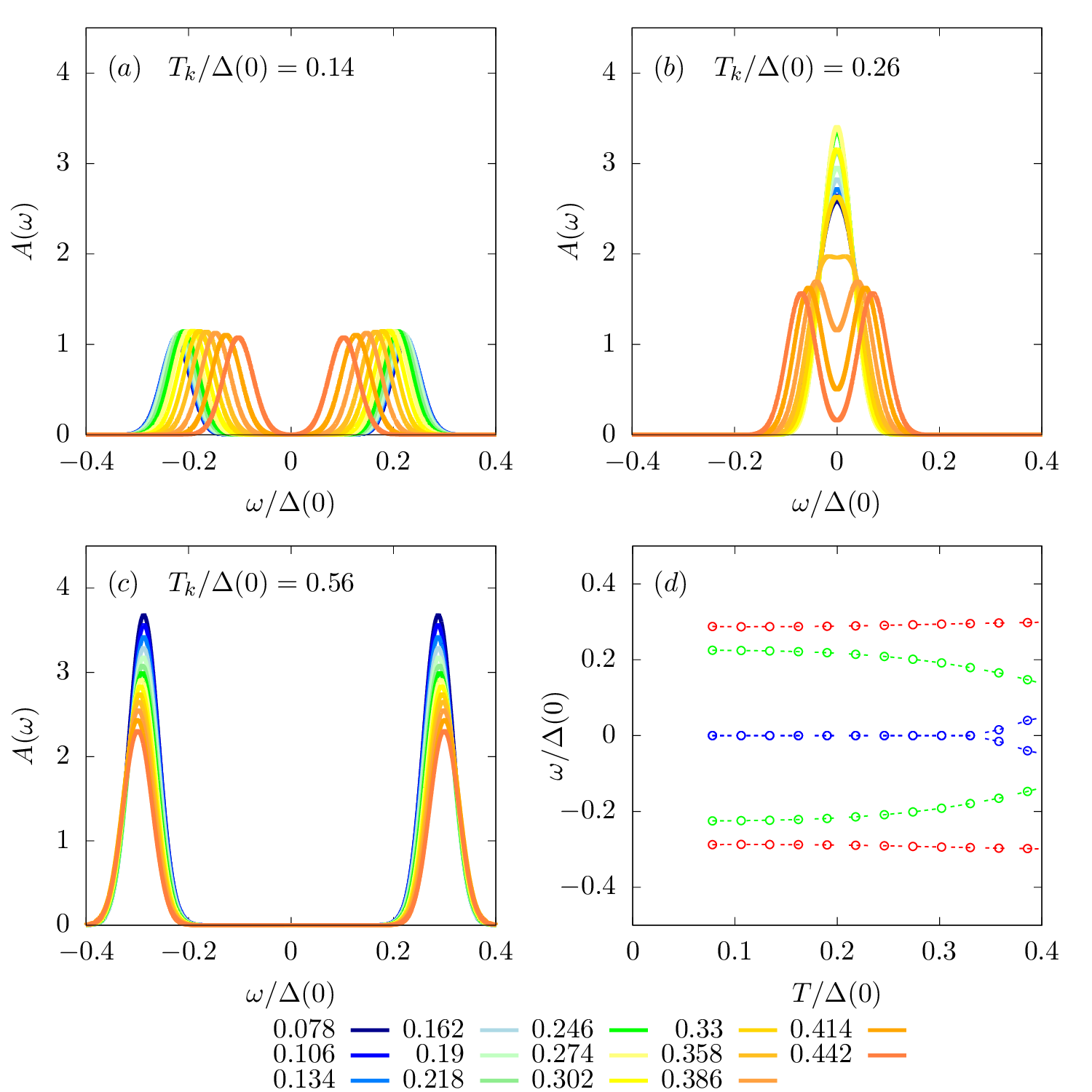}
\caption{\label{Fig2}(Color online) Spectral functions of the sub-gap states at finite temperatures.  We varied $T/\Delta(0)$ in figures $(a)-(c)$ while keeping $(a)$ $J/D=0.35, T_k/\Delta(0)=0.14$, $(b)$ $J/D=0.39, T_k/\Delta(0)=0.26$ and $(c)$ $J/D=0.45, T_k/\Delta(0)=0.56$.  In $(d)$,The YSR peak-positions obtained from Figs. (a), (b) and (c) vs $T/\Delta(0)$. The green cycles corresponding to the peak positions in $(a)$ if the $T$-dependence being carried out, blue cycles representing the peak positions in $(b)$ and the red cycles labelling the peak positions in $(c)$. The keys on the bottom dictating the values of $T/\Delta(0)$.}
\end{figure*}

\textsl{Numerical results ---} In Section A of the Supplement Material (SM),  we discuss how the energies of the ground and the first excited states of the Hamiltonian Eq. (\ref{eq1}) are calculated by NRG. The energy evolutions of the doublet and singlet states of the spin at the impurity site are obtained as functions of Kondo coupling $J$ at $T=0$.  This result is shown in Fig.S1.  There we rescale the energy value by subtracting the ground state $E_0$ at $J=0$.  In the weak coupling region such as $J/D< J_c/D\approx0.39$ which corresponding to $T_k/\Delta(0)=0.26$, the impurity spin ($S=1/2$) could not pair with any conduction electron, and thus the ground state has doublet degeneracy.  For $J>J_c$, it appears that the impurity spin can capture an electron from a Cooper pair and form a singlet ground state. This result is consistent with a previous calculation of \cite{doi:10.1143/JPSJ.62.3181}.  However, as to whether this "captured electron" is at the impurity site or not, so far has not been investigated. We argue from the feature of the spectral function at the impurity site, and this issue can be answered.

One of the primary efforts here is to obtain the temperature dependence of the spectral function that corresponds to the imaginary part of the $T$-matrix. This type of calculations has been performed by the NRG method\cite{ RevModPhys.47.773, doi:10.1143/JPSJ.61.3239, doi:10.1143/JPSJ.62.3181, RevModPhys.80.395, NRG}.  The method and the definition of the $T$-matrix are described in Section B of the Supplement Material.  In Section B of the SM, we also report the spectral function(see Fig.S2) of a Kondo impurity at $T=0$ in the presence of a magnetic field $h$ without SC in Fig.S2(a). The spectral function exhibits the Kondo resonance at zero energy for weak $h$, and the resonant peak will split into two as $g\mu_Bh/T_k^* > 0.51$ with $g=2$ and $\mu_B$ as the Bohr magneton. From Fig.S2(b), the transition from a single Kondo resonance to double resonances as $h$ varies appears to be of the first order. These results are consistent with those of Costi\cite{PhysRevLett.85.1504}.

In the presence of SC and a Kondo impurity, it is well known that the double peaks of YSR states in the spectral function at the impurity site are $\delta$-function like, and are located symmetrically with respect to the center of the SC gap. We plot the positions of the YSR peaks as functions of  $T_k/\Delta(0)$ at zero $T$ in Fig.S4 (see Section B of the SM). In the region of $0.9>T_k/\Delta(0) > 0.26$, the spin due to the impurity is in the singlet state or carries no net magnetic moment. We wish to understand why the in-gap YSR states, which is an essential feature of a magnetic impurity, still exist while the impurity paired with another electron to form the singlet-spin or nonmagnetic state. We argue that when $T_k$ is not too much larger than $\Delta(0)$, the impurity spin may loosely pair up with an electron from a Cooper pair to form a singlet. The ``paired electron" is not at the impurity site, and locally the impurity spin still retains its spin-doublet behavior and generates YSR states. However, for $T_k/\Delta(0) >1$ as shown in the inset of Fig.S4, the two YSR states separately move away from the mid part of the gap and toward the coherent peaks or edges of the gap as $T_k/\Delta(0)$ increases. When $T_k/\Delta(0) =5.2$(or $J/D=0.8$), the YSR peaks at $T=0$ are approaching to  the coherent peaks of the SC gap.  For  $J/D=1$,  we show that there exist no YSR states inside the gap.  In this limit, the pairing electron should be tightly bounded to the impurity site, and the spin state at the impurity site becomes a Kondo singlet which behaves like a nonmagnetic impurity\cite{PhysicsPhysiqueFizika.3.17}.

The square of the impurity magnetic moment $\langle M^2 \rangle$ as functions  of $T/\Delta(0)$ are calculated for several different values of  $T_k/\Delta(0)$ in Fig.\ref{Fig1}, here $M$ is the magnetic moment $M=S_{imp}^z$ due to the impurity defined in Section C of the SM.  The curves here show that for $T_k/\Delta(0)< 0.26$ and $> 0.26$, $<M^2>$ respectively equal to 0.25 (or $S_{imp}^z=1/2$, a doublet state), and 0 (or $S_{imp}^z=0$, a singlet state) at $T/\Delta(0) <10^{-2}$.  For  $T/\Delta(0) >10^{-2}$, the impurity spin could only be partially screened by the thermally excited electrons so that $<M^2>$ is always less than 0.25. But at $T>>T_k$, we expect $<M^2>$ should approach to 0.25 and the spin of the impurity becomes a doublet.  The insert figure showing the variation of  $<M^2>$ as a function of $T_k/\Delta(0)$ at $T=0$ is consistent with the those in Fig.S1 and Fig.S4 at $T=0$.

It appears that there exists a doublet to singlet transition at $T_k/\Delta(0) \approx 0.26$ for the spin state due to the Kondo impurity at $T=0$.   Let us now examine the spectral functions against $\omega/\Delta(0)$ at the impurity site for three different values of $T_k/\Delta(0)$ and several different temperatures. Here $\omega$ measures the bias energy.  The results are presented in Fig.\ref{Fig2}(a), (b) and (c). As one can see that as $T$ raises from zero, all the widths of the YSR peaks become broadened.  In Fig.\ref{Fig2}(a) with  $T_k/\Delta(0)=0.14$ (or $J/D=0.35$), the impurity spin state is in doublet at $T=0$ and the distance between the double YSR peaks shrinks as $T$ increases. In Fig.\ref{Fig2}(c) with $T_k/\Delta(0)=0.56$ (or $J/D=0.45$), the impurity spin state is a singlet at $T=0$ and the separation between the double YSR peaks is slightly expanding. One can also look into the spectral function shown in Fig.\ref{Fig2}(b) at the critical transition point with $T_k/\Delta(0) \approx 0.26$ (or $J/D=0.39$).  In this case, the double YSR peaks collapse into a single peak at $T=0$ which could split into two again as $T/\Delta(0)>0.35$. All these features imply that the critical transition point $T_k/\Delta(0)=0.26$ at $T=0$ should move to lower values at finite $T$.  In Fig.\ref{Fig2}(d), we plot the positions of YSR peaks shown in Figs.\ref{Fig2}(a), (b) and (c) against $T/\Delta(0)$.  The YSR states in the spectral function vs. the bias energy $\omega$ at the impurity site can be easily measured by the scanning tunneling microscopy (STM) experiments. It would be interesting to determine the spin state of the impurity at a very low temperature. This can be accomplished for a Kondo impurity in metal by measuring its magnetic susceptibility. However, in an SC, the magnetic susceptibility of the impurity cannot be detected, then the finite- $T$ behaviors exhibited in Fig.\ref{Fig2} can provide an unambiguous way to determine the spin state due to the magnetic impurity.  For instance, if the distance between the YSR peaks is shrinking as $T$ increases, the spin state at $T=0$ is a doublet, and if it is slightly increasing, then the impurity spin state should be a singlet.  The above conclusion is only valid when the SC gap $\Delta(T)$ decrease slightly as $T$ is raised from very low $T$ to a higher temperature $T<2/3T_c$ which should be true for the SC gap in BCS theory, here $T_c$ is the SC transition temperature.  As $T$ approaches to $T_c$, the separation between the YSR peaks would always be decreasing with $\Delta(T)$ regardless the value of $T_k/\Delta(0)$.

\begin{figure}
\centering
\includegraphics[scale=0.9]{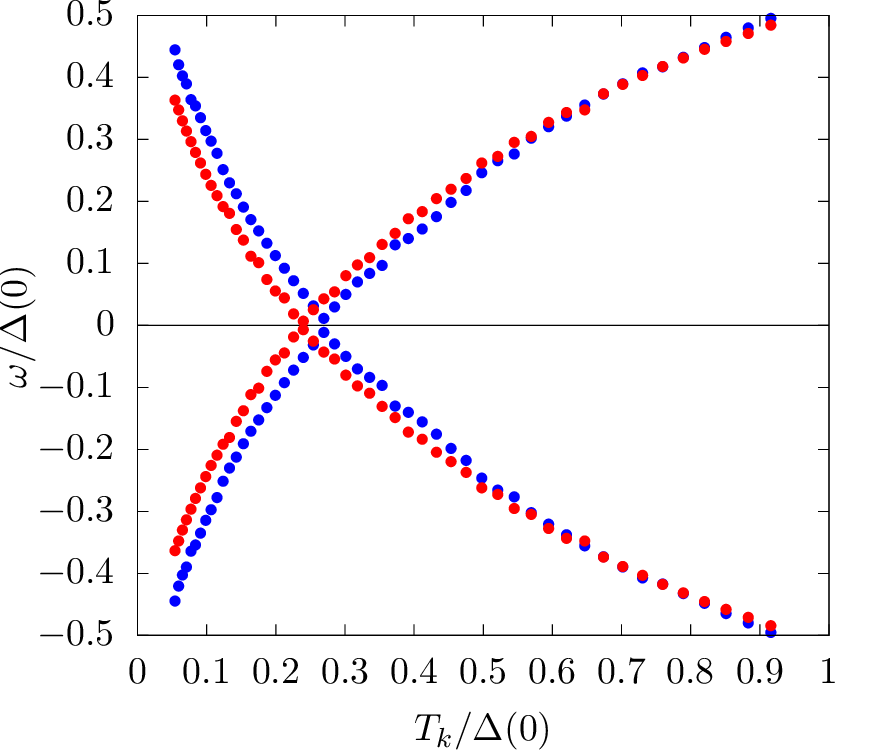}
\caption{\label{Fig3}(Color online) The sub-gap YSR-peak positions vs $T_k/\Delta(0)$ at $T/ \Delta(0)=0$ (blue dots ) and $T/\Delta(0)=0.358$ (red dots).}
\end{figure}

\begin{figure}
\centering
\includegraphics[scale=0.9]{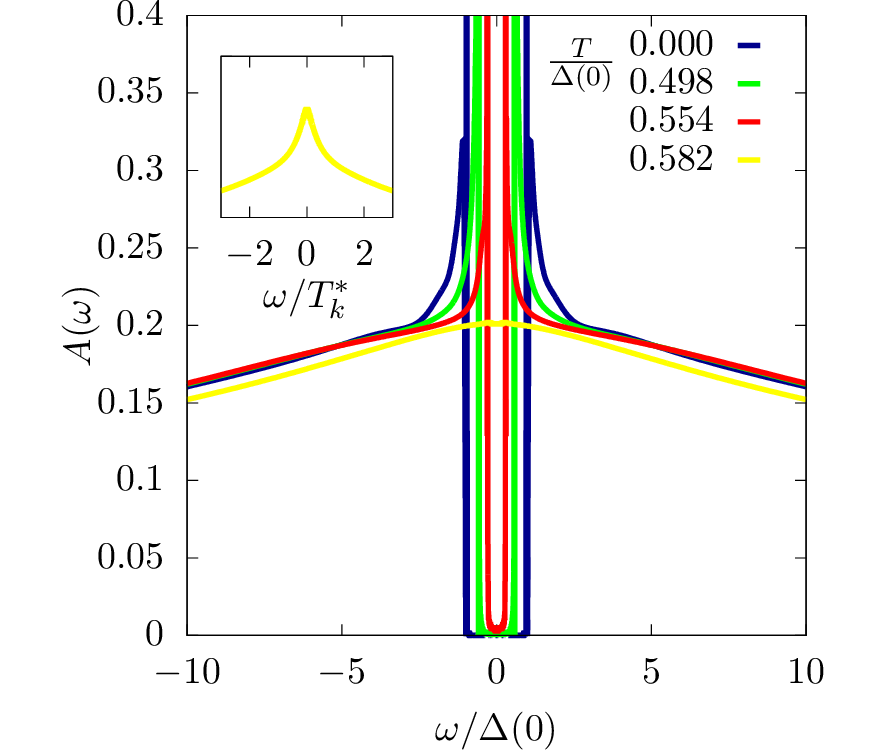}
\caption{\label{Fig4}(Color online) Spectral functions at $\Delta(T), T_k/\Delta(0)=5.2$ ($J/D=0.8$) and several different $T$.  The inset figure is the rescaled curve by $\omega/T_k^*$ . }
\end{figure}

The $T$-dependent behavior of the YSR peak positions is shown to originate from the $T$-dependent SC gap $\Delta(T)$.  The broadening of the YSR peaks is due to the thermally excited quasiparticles. But if one fixes $\Delta(T)=\Delta(0)$ as a $T$-independent quantity, then the peak positions would not be changed with $T$ as previous work has demonstrated for an Anderson impurity\cite{PhysRevB.93.195125}.  We are also able to obtain the same behavior for a Kondo impurity by setting $\Delta (T)=\Delta (0)$. In the present work, however, we set $\Delta (T)$ as the BCS SC gap at finite $T$ which has the expression as shown in Eq. (S13) in the SM.

For a better understanding what has been done in Fig.\ref{Fig2}, we plot the positions of YSR peaks from the spectrum function vs $T_k/\Delta(0)$ in Fig.\ref{Fig3}.  The curve with blue dots obtained at $T=0$ is identical to that in Fig.S4, and the curve with red dots is calculated for $T/\Delta(0)=0.358$. This result clearly indicates that the critical point $T_k/ \Delta(0) \approx 0.26$ at zero temperature is moving to the weaker Kondo coupling region at finite $T$.

So far we have studied the spectrum function only for moderate strength of $T_k/ \Delta(0)$ ($<0.9$). For an impurity with strong Kondo coupling such as being shown in the inset figure of Fig.S4 with $T_k/ \Delta(0)>5.0$, the YSR peaks move toward the edges or merge with coherent peaks of the SC gap. In Fig.\ref{Fig4}, we present the spectral function at the impurity site with $T_k/ \Delta(0)=5.2$  or$J/D=0.8$ for several different values of $T$. It can be seen that the YSR peaks are very close to the coherent peaks at $T/ \Delta(0)=0$.  At finite $T$, the two YSR peaks become broadened and merged completely with the coherent peaks at not too low temperatures. It appears that there are no longer YSR states inside the SC gap.   For $T/\Delta(0)=0.582 >T_c/\Delta(0)=0.57$, the SC no longer exists in the system and a broad peak (the orange curve) centered at $\omega=0$ shows up. If we re-plot the orange curve by a different energy scale $\omega/\ T_k^*$ as shown in the inset which exhibits the Kondo resonance at finite temperature without the SC. We also numerically calculated the integrated weight of the YSR peaks as a function of $J/D$ at $T=0$, the result is shown in Fig.S5 in Section B of SM. It is demonstrated there that as $J/D$ approaches 0 and 1, the integrated weight of the YSR peaks goes to 0. The maximum integrated weight comes around $J/D=0.5$. For $J=0$, there is no Kondo impurity, and there are no YRS peaks. For $J/D=1$, the YSR peaks are at the coherent peaks but with zero integrated weight, and that is the typical characteristics of a nonmagnetic impurity in which the impurity spin paired strongly with the spin of a conduction electron at the impurity site to form a rigid singlet state. The behavior for $J/D$ close to 1 is also consistent with the result for an Anderson impurity \cite{Bauer_2007}.

\textsl{Conclusion ---} We have studied the evolutions of the ground state and first excited energies of the Kondo Hamiltonian with SC at $T=0$ as functions of $T_k/\Delta(0)$. There the ground state is a doublet for $T_k/\Delta(0)<0.26$ and singlet for $T_k/\Delta(0)>0.26$. On the other hand, the spin at the impurity site is always in doublet state unless it can pair with an electron at the impurity site to form a singlet state in the region of $T_k/\Delta(0)>5.0$.  To determine whether the sample under experimental measurements is in doublet or singlet spin state at $T=0$ is an important issue to address.    We also show that the separation between the double YSR peaks in the spectral function decreases as $T$ is raised for $T_k/\Delta(0)<0.26$ while it is slightly increasing as $T$ is raised up to $T< 2/3T_c$ for $T_k/\Delta(0)>0.26$. This feature should be measurable by STM experiments, and the result could be used to unambiguously determine the spin state of the system at $T=0$.  The $T$-dependent spectral function for a strong Kondo impurity with $T_k/\Delta(0)=5.2$ or $J/D=0.8$  has also been calculated, and we find that the  YSR peaks are very close to the edges of the SC gap at $T=0$.  For $J/D=1$, we demonstrate that the YSR peaks are at the coherent peaks of the SC  but with zero weight which is the characteristics of a nonmagnetic impurity with singlet spin state\cite{PhysicsPhysiqueFizika.3.17}.

\textsl{Acknowledgments ---}  We thank for the help from Dr. R, {\v{Z}}itko and the useful discussions of Dr. Jian-Xin Zhu. Work at Houston is supported by the Robert A. Welch Foundation under the grant no. E-1146, and Texas Center for Superconductivity at the University of Houston. Work at Fudan University is by the National Key Research and Development Program of China (Grants No. 2017YFA0304204 and No. 2016YFA0300504), the National Natural Science Foundation of China (Grants No. 11625416 and No. 11474064), and the Shanghai Municipal Government under the Grant No. 19XD1400700.

\bibliography{manuscript}
\end{document}